\documentclass[aps,prc,twocolumn,superscriptaddress,showpacs]{revtex4}
\usepackage{graphicx}
\usepackage{longtable}

\newcommand{\fb}{$^\ast$}

\begin{document}

\title{Branchings in the $\gamma$ process path revisited}

\author{Thomas Rauscher}

\affiliation{Departement f\"ur Physik und Astronomie, Universit\"at Basel, 
CH-4056 Basel, Switzerland}

\email[]{Thomas.Rauscher@unibas.ch}
\homepage[]{http://nucastro.org}

\date{September 15, 2005}

\begin{abstract}
\noindent
The location of the ($\gamma$,p)/($\gamma$,n) and
($\gamma$,$\alpha$)/($\gamma$,n) line at $\gamma$-process temperatures
is discussed, using updated reaction rates based on global
Hauser-Feshbach calculations. The results can directly be
compared to classic $\gamma$-process
 discussions. The nuclei exhibiting the largest sensitivity to
uncertainties in nuclear structure and reaction parameters are
specified and suggestions for experiments are made. Additionally, the impact
of employing two recent global $\alpha$+nucleus
potentials is discussed. It is found that branchings at higher mass
depend more sensitively on these potentials.
The case of $^{146}$Sm/$^{144}$Sm production is addressed separately.
Also in this case, the more recent $\alpha$+nucleus potentials seem to
improve the issues concerning the production of these Sm isotopes in
massive stars.
In conclusion it is found that it is unlikely that the calculated
underproduction of $p$ nuclides in the Mo-Ru region is due to nuclear
physics deficiencies but that problems at higher mass number may still be
cured by improved nuclear input.
\end{abstract}

% 26.20.+f Hydrostatic stellar nucleosynthesis
% 26.30.+k Nucleosynthesis in novae, supernovae and other explosive environments
% 24.50.+g Direct reactions
%  25.40.-h Nucleon-induced reactions
% 25.40.Lw Radiative capture

\maketitle

\section{Introduction}
A number of proton-rich isotopes of naturally occurring stable nuclei cannot be
produced by neutron captures along the line of stability. They are called
$p$ isotopes. The currently most
favored production mechanism for those 35 $p$ isotopes
between Se and Hg is photodisintegration of intermediate
and heavy elements at high temperatures in late evolution stages of
massive stars, the so-called $\gamma$ process \cite{wh,rayet}.
However, not all $p$ nuclides can be produced satisfactorily, yet. A well-known
deficiency in the model is the underproduction of the Mo-Ru region but also
the region $151\leq A \leq 167$ is
underproduced, even in recent calculations \cite{rau02}.
It is not yet clear whether those deficiencies are due to the
astrophysical modelling or the employed nuclear physics.
Recent investigations have shown that
there still are considerable uncertainties in the description of nuclear
properties governing the relevant photodisintegration rates. This has
triggered a
number of experimental efforts to directly or indirectly determine reaction
rates
and nuclear properties for the $\gamma$ process (see, e.g.,
\cite{arngor,somorj,gyuri,fueloep,pdesc} and references therein). However, many such
investigations
focussed on nuclei in the $\gamma$-process path without considering whether the
rates involving these nuclei actually exhibit large uncertainties. In this work
the sensitivity of the location of the $\gamma$-process path on reaction rates
is investigated, showing which nuclei should be preferred in experimental
studies. 

A full $\gamma$-process network for a time-dependent calculation
comprises several hundreds to thousands of reactions. However, only
comparatively few reactions are actually relevant for the determination
of the reaction flow. Thus, an investigation of the involved nuclear
uncertainties can even be performed without relying on a full network
calculation but rather by studying ratios of photodisintegration rates which
determine how far the reaction path can extend to the proton-rich side
within an isotopic chain. In fact, such a ``model-free'' approach is not
limited to a given scenario, including seed nuclei and density profiles, but
has the advantage that principal limiting factors applying to any scenario
are derived. Concerning the astrophysical modelling, only a range of
temperatures has to be assumed but that can easily be extended. Here, I
show results for the ``classical'' range of $2.0\leq T_9 \leq 3.0$ (with $T_9$
being the temperature in 10$^9$ K).

\section{Branchings in the photodisintegration path}
\subsection{Definitions}
The $\gamma$ process starts with the photodisintegration of stable seed nuclei
which are present in the stellar plasma. The temperatures required for
significant photodisintegration can mostly only be achieved in explosive
burning, such as explosive O/Ne burning in massive stars. However, a recent
study also found some $\gamma$ processing already happening in late
evolution stages of massive stars before the actual explosion \cite{rau02}.
During the photodisintegration period, neutron, proton, and $\alpha$ emission
channels compete among each other and with $\beta^+$ decays further off
stability. In general, the nuclide destruction will commence with a
sequence of ($\gamma$,n) reactions, moving the abundances to the proton-rich
side. At some point in a chain of isotopes, ($\gamma$,p) and/or 
($\gamma$,$\alpha$) reactions will become faster than the neutron emission and
the flow will branch and feed another isotopic chain. At late times,
with decreasing temperature,
the photodisintegrations become less effective, leading to a shift of the
branch points and a take-over of $\beta^+$ decay. At the end of the process,
photodisintegrations cease quickly and the remaining unstable nuclei will
decay back to stability. Thus, the branchings established by the dominance
of proton and/or $\alpha$ emission over neutron emission are crucial in
determining the radioactive progenitors of the stable $p$ nuclei. The absolute
values of the rates determine the dynamics and time-scales which also
depend on the time-dependent temperature profile and thus on the chosen
astrophysical scenario. The branchings themselves only depend on the
ratios of the involved reaction rates.

Following the definition in \cite{wh}, a branch point is located at the
nucleus for which the condition $\lambda_{\gamma\mathrm{p}}+\lambda_{\gamma
\alpha}>\lambda_{\gamma \mathrm{n}}$ is
fulfilled for the first time when following an isotopic chain towards
decreasing neutron number $N$. The quantities $\lambda$ denote the
number of photodisintegrations per unit of time. For a reaction
$\gamma + \mathcal{A} \longrightarrow x+\mathcal{B}$, they are obtained by folding
the stellar photodisintegration cross section $\sigma_{\mathcal{A}(\gamma x)}^*$
of nucleus $\mathcal{A}$
with the energy distribution of the photons in the stellar photon gas
with temperature $T$:
\begin{equation}
\lambda_{\gamma x}=\frac{1}{\pi^2c^2\hbar^3}\int_0^{\infty}
\frac{\sigma_{\mathcal{A}(\gamma x)}^*\left(E_\gamma \right)E_\gamma^2}{e^{E_\gamma /kT}-1}
\, dE_\gamma \quad .
\end{equation}
The photodisintegration rate of nucleus $\mathcal{A}$
is related to the capture rate of nucleus $\mathcal{B}$ by
\begin{eqnarray}
\label{eq:detbal}
\lambda_{\gamma x}&\propto & e^{-\frac{S_x}{kT}} \left<\sigma v
\right>^*_{\mathcal{B}(x\gamma)}\\
&\propto & e^{-\frac{S_x}{kT}} \frac{1}{(kT)^{3/2}} 
\int_0^\infty \sigma_{\mathcal{B}(x\gamma)}^*\left( E\right)Ee^{-\frac{E}{kT}}dE\quad ,
\end{eqnarray}
with $n_x n_\mathcal{B} \left<\sigma v \right>^*_{\mathcal{B}(x\gamma)}$ being the
stellar capture rate, i.e.\ captures on the thermally excited nucleus $\mathcal{B}$, and
$n_x$, $n_\mathcal{B}$ being the number density of the projectiles $x$
and the nuclei $\mathcal{B}$, respectively
(see Refs.\ \cite{fow74,rath00} for details). The separation energy of
the emitted particle $x$ in the photodisintegrated nucleus $\mathcal{A}$ is denoted by
$S_x$. It is equal to the reaction $Q$ value of the capture reaction on nucleus $\mathcal{B}$.

The relation between
the different particle emission channels is a complex one but some
general rules can be stated. Since this has already been discussed
extensively in Ref.\ \cite{wh}, I limit myself to a brief reminder.
Equation \ref{eq:detbal} shows an exponential dependence of the
photodisintegration rate on the separation energy or capture $Q$ value.
For neutrons, the capture rate varies slowly compared to the $Q$ values
within an isotopic chain. Therefore, the effectivity of neutron emission
is governed by the neutron separation energies and will decrease for
increasingly proton-rich nuclei. Similar considerations apply
for proton and $\alpha$ emission except that for emission of charged
particles an additional exponential dependence on the Coulomb barrier
enters the cross section. Therefore, for comparable separation energies,
neutron emission will occur fastest and proton emission will dominate
over $\alpha$ emission. Due to the evolution of the separation energies,
there will be a nucleus within each isotopic chain, for which 
charged particle emission occurs faster than neutron emission. This is
the branch point according to the definition given above.
Moreover, it is expected that ($\gamma$,p) branchings will occur more often in
the lower mass range considered here, whereas ($\gamma$,$\alpha$) branchings
will be found more frequently in the higher mass range, due to the
distribution of separation energies.

For our considerations,
it is not only important where the branchings are located at a given
temperature but also how sensitive they are to a variation in the 
photodisintegration rates.
For instance, when a ($\gamma$,p) reaction is faster than both
neutron and $\alpha$ emission rates by a factor of, say, 100,
a variation of either rate by a factor of 10 will not have much effect
and the branching can be called robust. On the other hand, when the rates
are of the same magnitude, a small variation in any rate might either
remove the branching or change its nature (from ($\gamma$,p) to
($\gamma$,$\alpha$) or vice versa). Granted that theoretical rates are
not incorrect by arbitrarily large factors, the experimental study of such
sensitive branchings should be given priority.

It has to be noted that
there will be -- if at all -- only few, neighboring nuclei exhibiting
comparable rates in two or three emission channels due to the dependence
on the separation energy described above. Therefore, the location of a
branching cannot shift far away from the original position. However, when
rates are comparable also the actual value of the cross sections
is important.
Cross sections of nuclei relevant for the $\gamma$-process can be
calculated with the statistical Hauser-Feshbach model because the level
densities at the effective excitation energies are sufficiently high to
average over resonances \cite{rau97}.
Thus, sensitive branchings will also dependent on the nuclear properties
entering the statistical model. Among those, the optical potentials for
charged particle transmission will be the most important ones,
especially when dealing with projectile energies close to the Coulomb
barrier as it is the case for the $\gamma$ process.

\subsection{Updated branchings}
Let us start studying the branchings with modern rates
by applying a rate set (the set called FRDM
of \cite{rath00}) used in many stellar models, also the one 
of Ref.\ \cite{rau02}. Here, it will be called rate set \textbf{A}.
The rates were calculated using the NON-SMOKER
Hauser-Feshbach code and making use of a microscopic optical potential
for neutrons and protons \cite{jlm}. The global potential of \cite{McF}
was used for the $\alpha$ transitions. Further details of the code and
the inputs are described in \cite{rath00}.

Similar to Table 2 in \cite{wh} for $T_9=2.5$, the
branch points in the photodisintegration path appearing in the new
calculation are shown in the second, third, and fourth column of
Table \ref{tab:branch},
for three temperatures $T_9=2.0$, 2.5, 3.0. In this table
the neutron number $N$ of the branch point
is specified for each element.
The branching type is indicated by subscripts.
It can immediately be seen that branchings involving proton emission are
more important in the lower half of the mass range whereas $\alpha$ branch
points comprise most of the branchings in the upper mass range.
 
\begin{table*}
\caption{\label{tab:branch}Branch point nuclei obtained with three different
rate sets \textbf{A}, \textbf{B}, \textbf{C}; all rate sets were calculated with NON-SMOKER \cite{rath00} using different optical potentials for
$\alpha$ transmission: set \textbf{A} employs the potential of \cite{McF}, set \textbf{B} the one of \cite{avri}, and
set \textbf{C} the one of \cite{frohdip,raufroh}. Branchings of sets \textbf{B} and \textbf{C} differing from the standard branchings of
\cite{rath00} (rate set \textbf{A}) are marked by an
asterisk.}
\begin{ruledtabular}
\begin{tabular}{rr@{\hspace{20pt}}lll@{\hspace{20pt}}lll@{\hspace{20pt}}lll}
&&\multicolumn{9}{c}{Neutron number $N$ of branch point at given temperature $T_9$}\\
&&\multicolumn{3}{l}{Rate set \textbf{A}}&\multicolumn{3}{l}{Rate set \textbf{B}}&\multicolumn{3}{l}{Rate set \textbf{C}}\\
\multicolumn{1}{c}{Element}&\multicolumn{1}{l}{$Z$}&\multicolumn{1}{l}{2.0}&\multicolumn{1}{l}{2.5}&\multicolumn{1}{l}{3.0}&\multicolumn{1}{l}{2.0}&\multicolumn{1}{l}{2.5}&\multicolumn{1}{l}{3.0}&\multicolumn{1}{l}{2.0}&\multicolumn{1}{l}{2.5}&\multicolumn{1}{l}{3.0}\\
\hline
Se &  34 & 40$_{\alpha}$ &  40$_{\alpha}$ &  40$_{p,\alpha}$  &  40$_{\alpha}$ &  \fb40$_{p,\alpha}$ &  40$_{p,\alpha}$  &  \fb40$_{p,\alpha}$ &  \fb40$_{p,\alpha}$ &  \fb40$_{p}$  \\
Br &  35 & 46$_{p}$ &  44$_{p}$ &  44$_{p}$  &  \fb44$_p$ &  44$_{p}$ &  44$_{p}$  &  \fb44$_{p}$ &  44$_{p}$ &  44$_{p}$  \\
Kr &  36 & 44$_{p,\alpha}$ &  42$_{p}$ &  42$_{p}$  &  \fb42$_{p}$ &  42$_{p}$ &  42$_{p}$  &  \fb42$_{p}$ &  42$_{p}$ &  42$_{p}$  \\
Rb &  37 & 48$_{p}$ &  48$_{p}$ &  46$_{p}$  &  48$_{p}$ &  48$_{p}$ &  46$_{p}$  &  48$_{p}$ &  48$_{p}$ &  46$_{p}$  \\
Sr &  38 & 46$_{p}$ &  46$_{p}$ &  44$_{p}$  &  46$_{p}$ &  46$_{p}$ &  44$_{p}$  &  46$_{p}$ &  46$_{p}$ &  44$_{p}$  \\
 Y &  39 & 50$_{p}$ &  50$_{p}$ &  50$_{p}$  &  50$_{p}$ &  50$_{p}$ &  50$_{p}$  &  50$_{p}$ &  50$_{p}$ &  50$_{p}$  \\
Zr &  40 & 50$_{p}$ &  50$_{p}$ &  48$_{p}$  &  50$_{p}$ &  50$_{p}$ &  48$_{p}$  &  50$_{p}$ &  50$_{p}$ &  48$_{p}$  \\
Nb &  41 & 50$_{p}$ &  50$_{p}$ &  50$_{p}$  &  50$_{p}$ &  50$_{p}$ &  50$_{p}$  &  50$_{p}$ &  50$_{p}$ &  50$_{p}$  \\
Mo &  42 & 52$_{\alpha}$ &  50$_{p}$ &  50$_{p}$  &  52$_{\alpha}$ &  50$_{p}$ &  50$_{p}$  &  \fb50$_{p}$ &  50$_{p}$ &  50$_{p}$  \\
Tc &  43 & 54$_{p}$ &  52$_{p}$ &  52$_{p}$  &  54$_{p}$ &  \fb54$_{p}$ &  52$_{p}$  &  54$_{p}$ &  \fb54$_{p}$ &  52$_{p}$  \\
Ru &  44 & 54$_{\alpha}$ &  52$_{\alpha}$ &  52$_{p,\alpha}$  &  \fb52$_{\alpha}$ &  52$_{\alpha}$ &  \fb50$_{p}$  &  \fb52$_{\alpha}$ &  \fb52$_{p,\alpha}$ &  \fb50$_{p}$  \\
Rh &  45 & 56$_{p}$ &  56$_{p}$ &  56$_{p}$  &  56$_{p}$ &  56$_{p}$ &  56$_{p}$  &  56$_{p}$ &  56$_{p}$ &  56$_{p}$  \\
Pd &  46 & 56$_{\alpha}$ &  54$_{\alpha}$ &  54$_{p,\alpha}$  &  56$_{\alpha}$ &  \fb54$_{p,\alpha}$ &  \fb54$_{p}$  &  \fb54$_{p,\alpha}$ &  \fb54$_{p}$ &  \fb54$_{p}$  \\
Ag &  47 & 58$_{p}$ &  58$_{p}$ &  58$_{p}$  &  \fb60$_{p}$ &  58$_{p}$ &  58$_{p}$  &  \fb60$_{p}$ &  58$_{p}$ &  58$_{p}$  \\
Cd &  48 & 58$_{\alpha}$ &  58$_{\alpha}$ &  56$_{p}$  &  58$_{\alpha}$ &  \fb56$_{p}$ &  56$_{p}$  &  58$_{\alpha}$ &  \fb56$_{p}$ &  56$_{p}$  \\
In &  49 & 62$_{p}$ &  62$_{p}$ &  60$_{p}$  &  62$_{p}$ &  62$_{p}$ &  60$_{p}$  &  62$_{p}$ &  62$_{p}$ &  60$_{p}$  \\
Sn &  50 & 62$_{\alpha}$ &  60$_{p,\alpha}$ &  60$_{p}$  &  \fb60$_{p,\alpha}$ &  \fb60$_{p}$ &  60$_{p}$  &  \fb60$_{p,\alpha}$ &  \fb60$_{p}$ &  60$_{p}$  \\
Sb &  51 & 68$_{p}$ &  68$_{p}$ &  66$_{p}$  &  68$_{p}$ &  \fb66$_{p}$ &  66$_{p}$  &  68$_{p}$ &  \fb66$_{p}$ &  66$_{p}$  \\
Te &  52 & 68$_{\alpha}$ &  68$_{\alpha}$ &  66$_{\alpha}$  &  68$_{\alpha}$ &  \fb66$_{\alpha}$ &  66$_{\alpha}$  &  68$_{\alpha}$ &  \fb66$_{\alpha}$ &  \fb66$_{p,\alpha}$  \\
 I &  53 & 70$_{p}$ &  70$_{p}$ &  70$_{p}$  &  70$_{p}$ &  70$_{p}$ &  70$_{p}$  &  70$_{p}$ &  70$_{p}$ &  70$_{p}$  \\
Xe &  54 & 70$_{\alpha}$ &  68$_{\alpha}$ &  68$_{p,\alpha}$  &  70$_{\alpha}$ &  68$_{\alpha}$ &  68$_{p,\alpha}$  &  68$_{\alpha}$ &  \fb68$_{p,\alpha}$ &  \fb68$_{p}$  \\
Cs &  55 & 74$_{p}$ &  74$_{p}$ &  72$_{p}$  &  74$_{p}$ &  \fb72$_{p}$ &  72$_{p}$  &  74$_{p}$ &  \fb72$_{p}$ &  72$_{p}$  \\
Ba &  56 & 74$_{\alpha}$ &  72$_{\alpha}$ &  70$_{p,\alpha}$  &  \fb72$_{\alpha}$ &  \fb70$_{p,\alpha}$ &  \fb70$_{p}$  &  \fb72$_{\alpha}$ &  \fb70$_{p}$ &  \fb70$_{p}$  \\
La &  57 & 78$_{p}$ &  76$_{p}$ &  76$_{p}$  &  78$_{p}$ &  76$_{p}$ &  76$_{p}$  &  78$_{p}$ &  76$_{p}$ &  76$_{p}$  \\
Ce &  58 & 76$_{\alpha}$ &  74$_{\alpha}$ &  72$_{p,\alpha}$  &  76$_{\alpha}$ &  74$_{\alpha}$ &  \fb72$_{p}$  &  74$_{\alpha}$ &  \fb72$_{p}$ &  \fb72$_{p}$  \\
Pr &  59 & 80$_{p}$ &  80$_{p}$ &  80$_{p}$  &  80$_{p}$ &  80$_{p}$ &  \fb78$_{p}$  &  80$_{p}$ &  80$_{p}$ &  \fb78$_{p}$  \\
Nd &  60 & 78$_{\alpha}$ &  78$_{p,\alpha}$ &  74$_{p}$  &  \fb80$_{\alpha}$ &  \fb76$_{\alpha}$ &  74$_{p}$  &  \fb78$_{p,\alpha}$ &  \fb76$_{p,\alpha}$ &  74$_{p}$  \\
Pm &  61 & 84$_{\alpha}$ &  82$_{p}$ &  82$_{p}$  &  84$_{\alpha}$ &  82$_{p}$ &  82$_{p}$  &  \fb82$_{p}$ &  82$_{p}$ &  82$_{p}$  \\
Sm &  62 & 84$_{\alpha}$ &  80$_{p}$ &  80$_{p}$  &  84$_{\alpha}$ &  80$_{p}$ &  80$_{p}$  &  84$_{\alpha}$ &  80$_{p}$ &  80$_{p}$  \\
Eu &  63 & 88$_{\alpha}$ &  84$_{\alpha}$ &  82$_{p}$  &  \fb86$_{\alpha}$ &  84$_{\alpha}$ &  82$_{p}$  &  \fb84$_{\alpha}$ &  \fb82$_{p}$ &  82$_{p}$  \\
Gd &  64 & 88$_{\alpha}$ &  84$_{\alpha}$ &  82$_{p}$  &  88$_{\alpha}$ &  \fb86$_{\alpha}$ &  82$_{p}$  &  \fb86$_{\alpha}$ &  \fb82$_{p}$ &  82$_{p}$  \\
Tb &  65 & 88$_{\alpha}$ &  86$_{\alpha}$ &  84$_{p,\alpha}$  &  88$_{\alpha}$ &  \fb86$_{p,\alpha}$ &  84$_{p,\alpha}$  &  88$_{\alpha}$ &  \fb84$_{p}$ &  \fb84$_{p}$  \\
Dy &  66 & 90$_{\alpha}$ &  88$_{\alpha}$ &  86$_{\alpha}$  &  90$_{\alpha}$ &  88$_{\alpha}$ &  86$_{\alpha}$  &  90$_{\alpha}$ &  \fb86$_{\alpha}$ &  \fb84$_{\alpha}$  \\
Ho &  67 & 88$_{\alpha}$ &  88$_{p}$ &  88$_{p}$  &  \fb92$_{p,\alpha}$ &  88$_{p}$ &  88$_{p}$  &  \fb92$_{p}$ &  88$_{p}$ &  88$_{p}$  \\
Er &  68 & 92$_{\alpha}$ &  90$_{\alpha}$ &  88$_{\alpha}$  &  92$_{\alpha}$ &  90$_{\alpha}$ &  88$_{\alpha}$  &  92$_{\alpha}$ &  90$_{\alpha}$ &  88$_{\alpha}$  \\
Tm &  69 & 96$_{\alpha}$ &  92$_{p}$ &  92$_{p}$  &  96$_{\alpha}$ &  92$_{p}$ &  92$_{p}$  &  \fb94$_{p}$ &  92$_{p}$ &  92$_{p}$  \\
Yb &  70 & 96$_{\alpha}$ &  94$_{\alpha}$ &  92$_{p,\alpha}$  &  96$_{\alpha}$ &  94$_{\alpha}$ &  \fb92$_{\alpha}$  &  96$_{\alpha}$ &  \fb92$_{\alpha}$ &  \fb90$_{p,\alpha}$  \\
Lu &  71 & 96$_{\alpha}$ &  96$_{p}$ &  94$_{p}$  &  \fb98$_{\alpha}$ &  96$_{p}$ &  94$_{p}$  &  \fb96$_{p}$ &  96$_{p}$ &  94$_{p}$  \\
Hf &  72 & 100$_{\alpha}$ &  96$_{\alpha}$ &  94$_{\alpha}$  &  100$_{\alpha}$ &  96$_{\alpha}$ &  \fb94$_{p,\alpha}$  &  100$_{\alpha}$ &  \fb94$_{p,\alpha}$ &  \fb94$_{p}$  \\
Ta &  73 & 102$_{\alpha}$ &  98$_{p,\alpha}$ &  98$_{p}$  &  102$_{\alpha}$ &  98$_{p,\alpha}$ &  98$_{p}$  &  \fb100$_{\alpha}$ &  \fb98$_{p}$ &  98$_{p}$  \\
 W &  74 & 104$_{\alpha}$ & 102$_{\alpha}$ &  98$_{\alpha}$  &  104$_{\alpha}$ & \fb100$_{\alpha}$ &  98$_{\alpha}$  &  \fb102$_{\alpha}$ &  \fb98$_{\alpha}$ &  \fb96$_{p}$  \\
Re &  75 & 106$_{\alpha}$ & 102$_{\alpha}$ & 102$_{p,\alpha}$  &  \fb104$_{\alpha}$ & 102$_{\alpha}$ & \fb100$_{p,\alpha}$  &  \fb104$_{\alpha}$ & \fb102$_{p,\alpha}$ & \fb100$_{p}$  \\
Os &  76 & 106$_{\alpha}$ & 104$_{\alpha}$ & 102$_{\alpha}$  &  106$_{\alpha}$ & 104$_{\alpha}$ & \fb100$_{p}$  &  106$_{\alpha}$ & \fb102$_{\alpha}$ & \fb100$_{p}$  \\
Ir &  77 & 110$_{\alpha}$ & 106$_{\alpha}$ & 104$_{p}$  &  110$_{\alpha}$ & 106$_{\alpha}$ & \fb102$_{p}$  &  \fb108$_{\alpha}$ & \fb106$_{p,\alpha}$ & \fb102$_{p}$  \\
Pt &  78 & 109$_{\alpha}$ & 106$_{\alpha}$ & 106$_{\alpha}$  &  109$_{\alpha}$ & \fb108$_{\alpha}$ & \fb104$_{\alpha}$  &  \fb108$_{\alpha}$ & 106$_{\alpha}$ & \fb102$_{p,\alpha}$  \\
Au &  79 & 112$_{\alpha}$ & 110$_{\alpha}$ & 110$_{\alpha}$  &  112$_{\alpha}$ & 110$_{\alpha}$ & 110$_{\alpha}$  &  \fb110$_{\alpha}$ & 110$_{\alpha}$ & \fb108$_{p}$  \\
Hg &  80 & 110$_{\alpha}$ & 110$_{\alpha}$ & 108$_{\alpha}$  &  \fb112$_{\alpha}$ & 110$_{\alpha}$ & \fb106$_{\alpha}$  &  110$_{\alpha}$ & 110$_{\alpha}$ & \fb104$_{\alpha}$  \\
Tl &  81 & 112$_{\alpha}$ & 110$_{p}$ & 110$_{p}$  &  \fb112$_{p,\alpha}$ & 110$_{p}$ & 110$_{p}$  &  \fb112$_{p}$ & 110$_{p}$ & 110$_{p}$  \\
Pb &  82 & 114$_{\alpha}$ & 113$_{\alpha}$ & 110$_{\alpha}$
 &  114$_{\alpha}$ & \fb112$_{\alpha}$ & \fb112$_{\alpha}$  &  \fb113$_{\alpha}$ & \fb112$_{\alpha}$ & 110$_{\alpha}$
\end{tabular}
\end{ruledtabular}
\end{table*}

A direct comparison with Table 2 of \cite{wh} shows remarkable agreement
with a few exceptions. At first sight,
this is surprising insofar as the previous
rate predictions made use of a number of simplifying assumptions, such
as using equivalent square well potentials in the particle channels and
neglected excited states. However, the agreement can be
explained by the fact that the branch ratios are mainly
dependent on the $Q$ value ratios which are derived from experimental nuclear
masses. The aforementioned exceptions are Ba, W, Au, Hg
where the new branch points are shifted by 2 units to the more
neutron-rich side, Pb which is shifted by one unit, and Ce, Gd, Ho,
which have become more neutron-deficient by 2 neutrons. Only the
branching in Tl has been shifted by a larger amount, the branch point
has 4 neutrons less than previously. The branching type was modified
even less: a combined $\gamma \mathrm{p}+\gamma \alpha$ branching was
changed into a pure $\gamma \alpha$ one in Ba and Au, and a $\gamma
\mathrm{p}$ one has become a combined $\gamma \mathrm{p}+\gamma \alpha$
branching in Ta. (Combined branchings are nuclides at which both proton and
$\alpha$ emission is faster than neutron emission and within a factor of
3 of each other.) Incidentally, almost all altered branchings are within the
mass range $125\leq A\leq 150$ and $168\leq A\leq 200$ where
$\gamma$-process nucleosynthesis consistent with solar
$p$ abundances was found using the new rates \cite{rau02}, thus
underlining the improvement of the rate predictions.

\section{Experimental considerations}
Usually, experimental investigations primarily focus on nuclei close to
the branch points as given in Table \ref{tab:branch}.
However, they should rather focus on rates which
are sensitive to the nuclear input, i.e. nuclei for which
$\lambda_{\gamma \mathrm{n}}$, $\lambda_{\gamma \mathrm{p}}$, and
$\lambda_{\gamma \alpha}$ are close. To this end, Table \ref{tab:compact}
also shows the nuclei for which $\lambda_{\gamma \mathrm{p}}$ and
$\lambda_{\gamma \alpha}$ are within factors $f \leq 3$ and $f\leq 10$,
respectively, of
the $\lambda_{\gamma \mathrm{n}}$ rate. Subscripts indicate which rate
is close to $\lambda_{\gamma \mathrm{n}}$. Two subscripts indicate that
$\lambda_{\gamma \mathrm{p}}$ or $\lambda_{\gamma \alpha}$ are within
the quoted range but that they are also within a factor of 3 of each
other. The nuclei shown in Table \ref{tab:compact} were identified in
the NON-SMOKER calculations of \cite{rath00}, using the optical
$\alpha$+nucleus potential of \cite{McF}, similar to the results shown
for rate set \textbf{A} of Table \ref{tab:branch}.

\begingroup
\squeezetable
\begin{table}
\caption{\label{tab:compact}Nuclei with large rate uncertainties (derived
from rate set \textbf{A} \cite{rath00}, see
text); subscripts at each neutron number indicate which rate
($\lambda_{\gamma \mathrm{p}}$ or $\lambda_{\gamma \alpha}$) is close 
to the $\lambda_{\gamma \mathrm{n}}$ rate
within a factor of 3 and 10, respectively.}
\begin{ruledtabular}
\begin{tabular}{lp{2.5cm}p{2.5cm}p{2.5cm}p{2.5cm}}
&\multicolumn{3}{c}{Neutron number $N$ at given temperature $T_9$}\\
\multicolumn{1}{c}{$Z$}&\multicolumn{1}{l}{2.0}&\multicolumn{1}{l}{2.5}&\multicolumn{1}{l}{3.0}\\
\hline
 34 &   42$_{\alpha}$ &     &    \\
 35 &  46$_{p}$ &  46$_{p}$ &      \\
 36 &  44$_{p,\alpha}$ &  44$_{p}$ &    \\
 37 &     &  48$_{p}$ &  45$_{p}$, 48$_{p}$ \\
 38 &  43$_{p}$ &  43$_{p}$, 46$_{p}$ &  46$_{p}$ \\
 39 &  49$_{p}$ &  49$_{p}$ &  49$_{p}$ \\
 40 &  47$_{p}$ &  50$_{p}$ &  50$_{p}$ \\
 41 &    $_{ }$ &  46$_{p}$ &     \\
 42 &  52$_{\alpha}$ &    52$_{\alpha}$ &   \\
 43 &    $_{ }$ &  54$_{p}$ &   \\
 44 &  51$_{p}$, 54$_{\alpha}$ &  51$_{p}$, 52$_{\alpha}$ &  52$_{p,\alpha}$ \\
 46 &  53$_{\alpha}$, 56$_{\alpha}$ &  53$_{\alpha}$, 56$_{\alpha}$ &  53$_{\alpha}$, 54$_{\alpha}$ \\
 47 &  57$_{p}$, 60$_{p}$ &    &    \\
 48 &    $_{ }$ &  55$_{p,\alpha}$, 58$_{\alpha}$ &  55$_{p}$, 54$_{\alpha}$, 58$_{\alpha}$ \\
 49 &    $_{ }$ &  59$_{p}$, 62$_{p}$ &  59$_{p}$, 62$_{p}$ \\
 50 &  59$_{p,\alpha}$, 62$_{\alpha}$ &    & \\
 51 &  62$_{\alpha}$ &  68$_{p}$ &  63$_{p}$, 68$_{p}$ \\
 52 &  65$_{\alpha}$, 70$_{\alpha}$ &    68$_{\alpha}$ &  63$_{\alpha}$, 68$_{\alpha}$ \\
 53 &  67$_{p}$ &  67$_{p}$ &    \\
 54 &  67$_{\alpha}$, 72$_{\alpha}$ &  70$_{\alpha}$ &  68$_{p,\alpha}$, 70$_{\alpha}$ \\
 55 &  71$_{p}$ &  74$_{p}$ &  74$_{p}$ \\
 56 &  69$_{\alpha}$ & 72$_{\alpha}$, 74$_{\alpha}$ &  72$_{p,\alpha}$ \\
 57 &  73$_{p}$, 78$_{p}$ &  73$_{p}$, 78$_{p}$ &  78$_{p}$ \\
 58 &  76$_{\alpha}$, 78$_{\alpha}$ &  74$_{\alpha}$, 76$_{\alpha}$ &  72$_{\alpha}$, 74$_{p,\alpha}$ \\
 59 &  77$_{p}$, 84$_{\alpha}$ &   &  75$_{p}$, 80$_{p}$ \\
 60 &  75$_{\alpha}$, 80$_{\alpha}$, 84$_{\alpha}$ &  73$_{\alpha}$, 75$_{\alpha}$, 78$_{p,\alpha}$ &  73$_{p}$, 76$_{p,\alpha}$, 78$_{p}$ \\
 61 &  81$_{p}$, 84$_{\alpha}$ &   &  79$_{p}$ \\
 62 &  79$_{\alpha}$, 82$_{\alpha}$, 86$_{\alpha}$ &  77$_{\alpha}$, 82$_{\alpha}$, 84$_{\alpha}$ & 77$_{\alpha}$ \\
 63 &  86$_{\alpha}$, 88$_{\alpha}$ &  84$_{\alpha}$ &  84$_{\alpha}$ \\
 64 &  85$_{\alpha}$, 88$_{\alpha}$ &  79$_{p}$, 81$_{p}$, 86$_{\alpha}$ &  77$_{\alpha}$, 79$_{p}$, 81$_{p}$, 85$_{\alpha}$ \\
 65 &  87$_{\alpha}$, 90$_{\alpha}$ &  86$_{\alpha}$, 88$_{\alpha}$ &  86$_{p,\alpha}$, 88$_{\alpha}$ \\
 66 &  83$_{\alpha}$, 87$_{\alpha}$, 90$_{\alpha}$ &  87$_{\alpha}$, 88$_{\alpha}$ &  85$_{\alpha}$, 86$_{\alpha}$, 88$_{\alpha}$ \\
 67 &  90$_{p,\alpha}$, 92$_{p,\alpha}$ &  83$_{p,\alpha}$, 87$_{p,\alpha}$ &  85$_{\alpha}$ \\
 68 &  89$_{\alpha}$, 91$_{\alpha}$, 94$_{\alpha}$ &  83$_{\alpha}$, 87$_{\alpha}$, 90$_{\alpha}$, 92$_{\alpha}$ &  83$_{p}$, 87$_{\alpha}$, 88$_{\alpha}$, 90$_{\alpha}$ \\
 69 &  89$_{\alpha}$, 91$_{\alpha}$, 96$_{\alpha}$ &  89$_{p}$, 94$_{p,\alpha}$ &  89$_{p}$, 94$_{p}$ \\
 70 & 91$_{\alpha}$, 93$_{\alpha}$, 98$_{\alpha}$ & 89$_{\alpha}$, 94$_{\alpha}$ &  87$_{\alpha}$, 89$_{\alpha}$, 92$_{p,\alpha}$, 94$_{\alpha}$ \\
 71 &  95$_{\alpha}$, 98$_{\alpha}$, 100$_{\alpha}$ &  93$_{p}$, 96$_{p}$ &  93$_{p}$, 96$_{p}$ \\
 72 & 95$_{\alpha}$, 100$_{\alpha}$, 102$_{\alpha}$ &  93$_{\alpha}$, 96$_{\alpha}$, 98$_{\alpha}$ &  89$_{\alpha}$, 91$_{\alpha}$, 94$_{\alpha}$, 96$_{\alpha}$ \\
 73 & 97$_{\alpha}$, 99$_{\alpha}$, 104$_{\alpha}$ &  95$_{p}$, 100$_{\alpha}$, 102$_{\alpha}$ &  95$_{p}$, 100$_{\alpha}$, 102$_{\alpha}$ \\
 74 &  99$_{\alpha}$, 101$_{\alpha}$, 104$_{\alpha}$ &  95$_{\alpha}$, 97$_{\alpha}$, 100$_{\alpha}$, 102$_{\alpha}$ &  93$_{p,\alpha}$, 95$_{\alpha}$, 98$_{\alpha}$, 100$_{\alpha}$ \\
 75 & 101$_{\alpha}$, 106$_{\alpha}$ &  99$_{p,\alpha}$, 104$_{\alpha}$ &  99$_{p}$, 102$_{\alpha}$ \\
 76 & 103$_{\alpha}$, 108$_{\alpha}$ & 99$_{\alpha}$, 101$_{\alpha}$, 104$_{\alpha}$, 106$_{\alpha}$ &  97$_{p,\alpha}$, 99$_{\alpha}$, 102$_{\alpha}$, 104$_{\alpha}$ \\
 77 & 103$_{\alpha}$, 105$_{\alpha}$, 110$_{\alpha}$ & 106$_{\alpha}$, 108$_{\alpha}$ & 106$_{p,\alpha}$  \\
 78 & 107$_{\alpha}$, 109$_{\alpha}$, 110$_{\alpha}$, 112$_{\alpha}$ & 103$_{\alpha}$, 105$_{\alpha}$, 108$_{\alpha}$ & 101$_{\alpha}$, 103$_{\alpha}$, 106$_{\alpha}$  \\
 79 & 111$_{\alpha}$, 112$_{\alpha}$ & 107$_{p,\alpha}$, 109$_{\alpha}$ & 105$_{\alpha}$, 107$_{p}$  \\
 80 &    & 107$_{\alpha}$, 109$_{\alpha}$, 110$_{\alpha}$ & 105$_{\alpha}$, 108$_{\alpha}$, 110$_{\alpha}$  \\
 81 & 112$_{\alpha}$ & 109$_{\alpha}$ &  \\
 82 & 111$_{\alpha}$, 118$_{\alpha}$ & 105$_{\alpha}$, 107$_{\alpha}$, 109$_{\alpha}$, 112$_{\alpha}$, 113$_{\alpha}$ & 107$_{p,\alpha}$, 109$_{\alpha}$, 110$_{\alpha}$, 113$_{\alpha}$
\end{tabular}
\end{ruledtabular}
\end{table}
\endgroup

The factors were chosen according to the assumed uncertainties in the
predicted rates. The $\gamma$ process path is not located very far from
stability, therefore a comparison of theory and experiment for stable
targets gives a good estimate of the involved uncertainties. For neutron
capture, an average uncertainty of 30\% was found \cite{rau97}.
Due to the Coulomb barrier, charged particle reactions are more sensitive
to the surface potentials. While many proton captures are theoretically
described with a similar accuracy as neutron captures, some local deviations
of up to factors $2-3$ have been found. By far the largest uncertainty is
found in reactions involving low-energy $\alpha$ particles (see, e.g.,
\cite{somorj}). The photodisintegration rates are expected to show
similar uncertainties as the capture rates, provided the $Q$ value is
known accurately. Consequently, ($\gamma$,p) rates are considered with a
variation by a
factor of 3 and ($\gamma$,$\alpha$) ones within a factor of 10. An extended
table also including ($\gamma$,p) uncertainties up to a factor of 10 can
be found in \cite{rau05nic}.

As pointed out above, experiments targeting the sensitive rates given in
Table \ref{tab:compact} will have direct impact on $\gamma$-process
nucleosynthesis. Among them, sensitive rates at branch points
(coinciding with the nuclei given in Table \ref{tab:branch}) will be the
most important. Because of the rapid evolution of $Q$ values within an
isotopic chain, reactions on nuclei next to branchings are usually not
important anymore.

Concerning the reaction type, channels with charged particles are more sensitive
than neutron emission. The latter plays a role in determining the
time scale when shifting isotopes from stability to the proton-rich side.
Due to the $Q$ value, ($\gamma$,n) reactions on targets with an even neutron
number are slower than the ones on odd-$N$ targets. Since the time scale in
a reaction chain is governed by the slowest rates, those have to be checked
primarily.

Recently, there has been increased interest in directly studying
photodisintegration reactions in experiments with Bremsstrahlung or Laser
inverse-Compton scattering photons, also
motivated by the astrophysical importance of such reactions \cite{utso}.
However, most of the
relevant $\gamma$ transitions cannot be accessed in this manner \cite{mohr}.
Therefore such
measurements can be used to test reaction models selectively but not to directly
access the required reaction for the $p$ process. This can be achieved by
measuring the capture reaction in the relevant energy range, from which
the reverse rate can straighforwardly be derived by applying detailed
balance \cite{rath00} when the $Q$ value is known to good accuracy. This
even applies in the case of reactions with negative $Q$ value for
capture because the stellar photodisintegration rate differs by several
orders of magnitude from the ground state photodisintegration rate
measured in the laboratory \cite{mohr}. In
consequence, the nuclei given in the tables are then the {\it final}
nuclei of the respective capture reactions.

Many of the sensitive branchings occur at nuclei with half-lives of less than
a month. Future radioactive ion beam facilities such as GSI (Germany) and RIKEN
(Japan) upgrades or the planned RIA (USA) will be able to access most of them
although it remains an open question
whether reaction studies can be performed. Conventional
nuclear experiments are limited to stable or long-lived targets.
An overview of the most important
reactions on stable or long-lived targets is presented in Table \ref{tab:exp}.
There may be data available for several of the given reactions but not
necessarily in the $p$ process energy range. Extrapolations into the
energy range are discouraged, especially for the lighter targets, because
of possible resonance contributions, neglected in statistical model
calculations. In Table \ref{tab:exp}, priority group 1
includes reactions in sensitive branchings, priority group 2 are reactions
which could become new branchings if their rate is found to be increased. 

Finally, it should be noted that in this ``model-free'' approach equal
weight is given
to each $Z$ chain. In an astrophysical network calculation, the impact of
certain isotopic chains may be enhanced or suppressed according to the
chosen seed abundance as more or less seed nuclei are available for
photodisintegration for a given element.
However, the main features will still be determined
by the underlying nuclear physics.
 
\begingroup
\squeezetable
\begin{table}
\caption{\label{tab:exp}Suggestions for reactions to be studied experimentally.
Shown are sensitive reactions involving stable or long-lived ($T_{1/2}\geq10^6$
a) targets. Unstable targets are marked by an asterisk, naturally occuring
unstable nuclides with superscript 'n'. Note that $\alpha$
capture on the unstable targets shown here always has a negative $Q$ value.}
\begin{ruledtabular}
\begin{tabular}{ll}
&\multicolumn{1}{c}{Target nuclei}\\
\hline
Priority 1:&\\
(p,$\gamma$)&$^{80}$Se, $^{79}$Br, $^{84}$Kr, $^{89}$Y, $^{93}$Nb, $^{97}$Tc$^*$,
$^{110}$Cd, $^{118}$Sn,\\
&$^{128}$Xe, $^{134}$Ba, $^{138}$Ce\\
($\alpha$,$\gamma$)&$^{76}$Se, $^{92}$Mo, $^{94}$Mo, $^{96}$Ru, $^{98}$Ru,
$^{102}$Pd, $^{108}$Cd, $^{116}$Sn,\\
&$^{124}$Xe, $^{130}$Ba,
$^{141}$Pr, $^{148}$Sm$^\mathrm{n}$, $^{152}$Gd$^\mathrm{n}$,
$^{150}$Gd$^\mathrm{n}$, $^{154}$Dy$^\mathrm{n}$,\\
&$^{168}$Yb, $^{174}$Hf$^\mathrm{n}$\\
\hline
Priority 2:&\\
(p,$\gamma$)&$^{96}$Mo, $^{106}$Pd, $^{150}$Gd$^*$,
$^{156}$Dy, $^{158}$Dy, $^{162}$Er\\
($\alpha$,$\gamma$)&$^{72}$Ge, $^{90}$Zr, $^{118}$Sn, $^{120}$Te, $^{122}$Te,
$^{126}$Xe, $^{132}$Ba, $^{139}$La,\\
&$^{136}$Ce, $^{140}$Ce,
$^{142}$Nd, $^{144}$Nd$^\mathrm{n}$, $^{146}$Sm$^*$, $^{151}$Eu,
$^{156}$Dy,\\
&$^{158}$Dy, $^{164}$Er, $^{170}$Yb,
$^{180}$W, $^{184}$Os, $^{186}$Os$^\mathrm{n}$, $^{196}$Hg
\end{tabular}
\end{ruledtabular}
\end{table}
\endgroup

\section{Different $\alpha$+nucleus potentials}
In recent investigations
it has become apparent that the most important problem for the
calculation of reaction rates 
is the determination of the optical $\alpha$+nucleus potentials at low energies
(see \cite{somorj,gyuri,fueloep,pdesc} and references therein).
Thus, the $\lambda_{\gamma \alpha}$ rates bear the largest inherent
uncertainty whereas $\lambda_{\gamma \mathrm{n}}$ and $\lambda_{\gamma
\mathrm{p}}$ have been found generally well predicted, with a few
exceptions \cite{gyuri,fueloep}, as mentioned above.

It is interesting to view the changes brought upon by using different optical
potentials. Rate set \textbf{A}, discussed so far,
has been calculated using the potential by
\cite{McF} which was fitted to $\alpha$ scattering data across a large mass
range at intermediate energies. Although the potential works well also for many reaction
data even at the comparatively low projectile energies of astrophysical
interest, large deviations have been found for a number of cases. This
motivated the quest for finding improved optical $\alpha$ potentials which
also work well at energies close to the Coulomb barrier.

From a number of
global potentials \cite{rau,demet,avri,frohdip,raufroh},
I choose two publicly available ones
for comparison here. The recent potential of \cite{avri}
has been fitted to a large data compilation at low and intermediate energies
and describes well both scattering and reaction data across a large
mass range (see, however, Ref.\ \cite{avrinew} for a possible necessity
for modifications).
Rate set \textbf{B} was calculated with this potential.

The potential of \cite{frohdip,raufroh} was employed for rate set
\textbf{C}. It is fitted to low-energy
reaction data around mass $A\simeq 145$. Although it does not describe
scattering data, reaction data at lower masses ($A>90$) are
reproduced well \cite{avrinew,galaviz}. In Ref.\ \cite{avrinew} it is argued
that optical potentials may depend on the nuclear temperature and thus
the idea is supported that this potential may be well suited to describe
reactions even though it is not suited for scattering data.

Columns six to eleven of Table \ref{tab:branch} show the
branchings obtained with the two potentials (all other inputs remained
unchanged). Branchings differing from the ones obtained with the standard
rate set \textbf{A} either by neutron number or by branch type are marked
by an asterisk. As expected, the branchings in the lower mass range remain
mostly unchanged whereas considerable changes are found for the
heaviest nuclides. With a few exceptions,
the branchings are shifted to lower neutron number within an
isotopic chain by about 2 units, i.e.\ further off stability. This helps
the faster processing of material to lower charge number and may indeed help
to cure the underproduction in the region between Eu and Yb found
in \cite{rau02}.

\section{Photodisintegration of $^{148}$Gd}
The ($\gamma$,n)-($\gamma$,$\alpha$) branching at $^{148}$Gd has received
frequent attention \cite{wh,somorj,woogd,raulett}. It determines the production
ratio of $^{144}$Sm and $^{146}$Sm (as decay product of
$^{146}$Gd($\beta^+$)$^{146}$Eu($\beta^+$)$^{146}$Sm). The abundance ratio
of $^{144}$Sm and $^{142}$Nd can be measured in circumstellar grains
embedded in meteorites \cite{pap}. Since $^{142}$Nd is a decay product of
the long-lived $^{146}$Sm ($T_{1/2}=1.03\times 10^8$ a), the ratio can be
either used as a chronometer if the initial production ratio is known or
to determine the initial production ratio if the time scale is known.

As can be seen in Table \ref{tab:branch},
this branching mainly acts at around $T_9=2.5$, producing $^{144}$Sm.
The nucleus $^{144}$Sm is also produced at higher temperature although
neutron emission dominates. It can still be reached via two
($\gamma$,p) branchings at $^{146}$Gd and $^{145}$Eu. In both cases,
$^{146}$Sm production is suppressed. At lower temperature,
$^{144}$Sm is bypassed because the reaction flow branches off already at
larger $N$ in both the Gd and Sm isotopic chains. Thus, the production
ratio of $^{144}$Sm and $^{146}$Sm is not only determined by the ratio
$\lambda_{\gamma \mathrm{n}}$/$\lambda_{\gamma \alpha}$ but also by the
temperature history. Therefore, a change in this rate ratio does not
linearly enter the final production ratio, as was also found in
\cite{somorj}.

Moreover, it has to be considered that with an improved $\alpha$+nucleus
potential not only the $\alpha$ emission of $^{148}$Gd will change but also
others in its vicinity. This effect can be seen in Table \ref{tab:branch}
in the results obtained with the other two global optical potentials
(rate sets \textbf{B} and \textbf{C}).
With the potential of \cite{avri}, the situation remains unchanged for the
high and the low temperature region. At intermediate temperature, an
$\alpha$ branching appears already at $^{150}$Gd, feeding into $^{146}$Sm.

An even larger change can be found when using the potential of
\cite{frohdip,raufroh}. Again, the situation remains similar to the standard
case at $T_9=3.0$. At $T_9=2.5$ the main branching in the chain
still is the proton branching at $^{146}$Gd, bypassing $^{146}$Sm.
This time also the branching at $T_9=2.0$ is shifted. It appears as an
$\alpha$ emission at $^{150}$Gd. Thus, $^{146}$Sm will only be produced
at low temperature but possibly at a higher level than found in the other two
calculations.

Although detailed production ratios can only be obtained in time-dependent
simulations, my estimate is that the production of $^{146}$Sm will be
enhanced with the recent global potentials. This appears to be a trend into
the desired direction as the predicted $^{146}$Sm/$^{144}$Sm production
ratios \cite{wh,rayet,raulett} were
too low compared to the values derived in \cite{pap}.

\section{Summary}
The nuclear uncertainties in the $\gamma$ process were explored and a number
of sensitive reaction rates were identified.  Some of the rates can be
studied experimentally. However, it became clear that nuclear uncertainties
cannot be the cause for the underproduction of $p$ nuclides in the Mo-Ru
region as the branchings seem to be robust. This appears to be consistent
with other considerations, e.g., it was pointed out already in \cite{wh}
that Mo and Ru would still remain underabundant even if all seed material
would be ideally photodisintegrated. Thus, a different production mechanism
has to be found, perhaps involving higher temperatures and/or a different
seed composition. On the other hand, the less robust $\alpha$ branchings
dominate in the higher mass range and further (experimental)
work has to be done to provide a sound footing of $\gamma$-process calculations
there. It is conceivable that the deficiences found in the $p$ mass range
$151\leq A\leq 167$ are due to nuclear uncertainties.

\begin{acknowledgments}
Thanks go to P. Mohr and Zs.\ F\"ul\"op for stimulating discussions.
This work was supported
by the Swiss National Science Foundation (grant 2000-105328/1).
\end{acknowledgments}

\end{document}